# Urban Form and Structure Explain Variability in Spatial Inequality of Property Flood Risk among US Counties

Junwei Ma[1], Ali Mostafavi[1*]

[1] Urban Resilience.AI Lab, Zachry Department of Civil and Environmental Engineering, Texas A&M University, College Station, United States.
[*] Corresponding author: Ali Mostafavi, amostafavi@civil.tamu.edu.

**ABSTRACT**

Understanding the relationship between urban form and structure and spatial variation of property flood risk has been a longstanding challenge in urban planning, flood risk management, and city emergency management. Yet limited data-driven insights exist regarding the extent to which variation in spatial inequality of property flood risk in cities can be explained by heterogenous features of urban form and structure. In this study, we explore eight key features (i.e., population density, point of interest density, road density, minority segregation, income segregation, urban centrality index, gross domestic product, and human mobility index) related to urban form and structure to explain variability in spatial inequality of property flood risk among 2567 US counties. Using rich datasets related to property flood risk, we quantify spatial inequality in property flood risk and delineate features of urban form and structure using high-resolution human mobility and facility distribution data. We identify significant variation in spatial inequality of property flood risk among US counties with coastline and metropolitan counties having the greatest spatial inequality of property flood risk. The results also reveal variations in spatial inequality of property flood risk can be effectively explained based on principal components of development density, economic activity, and centrality and segregation. Using a classification and regression tree model, we demonstrate how these principal components interact and form pathways that explain levels of spatial inequality in property flood risk in US counties. The findings offer important insights for the understanding of the complex interplay between urban form and structure and spatial inequality of property flood risk and have important implications for integrated urban design strategies to address property flood risk as cities continue to expand and develop.

**Introduction**

Climate change has increased the frequency and intensity of flooding, elevating concern about the threat it poses to millions of people [1-4], including extensive property damages [5-7]. While flood risk is often examined with a focus on natural factors, such as rainfall and land topography [8-10], the importance of urban form and structure in shaping the spatial distribution of flood risk in cities is increasingly being recognized [11, 12]. In particular, understanding the relationship between urban form and structure and spatial distribution of flood risk may hold the key to integrated urban design strategies to effectively address flood risk as cities continue to grow and develop.

Urban form and structure refer to the spatial configuration and organization of cities, such as development patterns, facility distribution, and economic activity [13-15]. Urban form and structure capture characteristics such as population distribution [16], development density [17], human mobility [18], and the centralization of infrastructure [19] that can shape the spatial distribution of flood risk. Yet data-driven insights are missing to inform about the relationship between urban form and structure and spatial distribution of flood risk. This limitation has hindered development and implementation of integrated urban design strategies to inform growth and development of cities while addressing flood risks to people and properties.

Of particular importance is uncovering the extent to which urban form and structure explain variation in spatial inequality of property flood risk in cities. Spatial inequality of property flood risk captures the extent to which properties located in different areas of a city have similar level of flood risk. The risk of property flooding is not distributed equally across all areas and communities in a city [20, 21]. Studies have shown that low-income communities and communities of color are often more vulnerable to flooding than wealthier, white communities [22-25]. Spatial inequality of property flood risk is in part influenced by the patterns of growth and development in cities [26]. A critical and yet unanswered question is the extent to which features of urban form and structure (such as development density, economic activity, and centrality and segregation) can explain the variation in spatial inequality of property flood risk among cities.

To address this research gap, this study investigates the extent to which urban form and structure explain variability in spatial inequality of property flood risk among US counties. The main research questions guiding the study are twofold: (1) What is the extent of spatial inequality in property flood risk in US counties?; (2) To what extent are spatial inequality of property flood risk explained by different features of urban form and structure?. To answer these questions, we use rich datasets related to property flood risk to quantify spatial inequality of property flood risk in US counties and delineate features of urban form and structure using high-resolution human mobility and facility distribution data. We begin by evaluating spatial inequality of property flood risk using the metric of spatial Gini index (SGI), a measure of spatial inequality, for 2567 counties in the United States, identifying significant variations in spatial inequality of property flood risk across counties. We then explore how urban form and structure may be shaping this spatial inequality of property flood risk, by examining eight distinct urban features (i.e., population density, point of interest density, road density, minority segregation. income segregation, urban centrality index, gross domestic product, and human mobility index) to assess their potential relationships. We use principal component analysis to identify three key factors shaping spatial inequality of property flood risk: development density, centrality and segregation, and economic activity. We then develop a classification and regression tree model to examine ways these factors interact in forming pathways leading to different levels of spatial inequality in property flood risk in US counties. Finally, we discuss integrated urban design strategies for addressing spatial inequality of property flood risk to inform future growth and expansion of cities. Our study provides unique and valuable insights into the intricate relationship between urban form, urban structure, and spatial inequality of property flood risk. These insights carry significant implications for integrated urban design strategies aimed at mitigating property flood risk, particularly as cities undergo further expansion and development.

## Results

### Variation in spatial inequality of property flood risk among US counties

To explore the variation in spatial inequality of property flood risk among US counties, the SGI was calculated based on the Flood Factor dataset from First Street Foundation (see Methods for detail). Probability density function (PDF) and complementary cumulative distribution function (CCDF) of the SGI were plotted in Fig 1 (a) and (b) to examine the distribution characteristics. The results show that the PDF has a unimodal distribution with a mean of 0.255 and a standard deviation of 0.135. The CCDF, on the other hand, exhibits a heavy-tailed distribution, indicating the existence of a number of US counties with significant spatial inequality of property flood risk. The 20% of counties with the highest SGI values account for 36% of the property flood risk, while the 20% of counties with the lowest SGI values only account for 7% of the property flood risk. These results suggest a significant variation in spatial inequality of property flood risk across US counties, highlighting the importance of examining the underlying factors shaping this disparity.

To compare the spatial distribution of SGI and overall flood risk, we created county-level visualizations of SGI and overall flood risk in the United States. The results show a significant variation in the SGI across US counties, with some counties exhibiting extremely high levels of spatial inequality in property flood risk (top panel of Fig 1 (c)). Interestingly, the high flood-risk property areas shown in the overall flood risk visualization (bottom panel of Fig 1 (c)), such as counties in the Gulf Coast region, do not necessarily correspond to high levels of spatial inequality in our analysis (see highlighted areas in Fig 1 (c)). On the contrary, some counties in California and the US Northeast have a much higher SGI than the counties in Texas, yet they have lower overall flood risk. The findings indicate that the spatial inequality of property flood risk is not solely determined by the extent of overall flood risk.

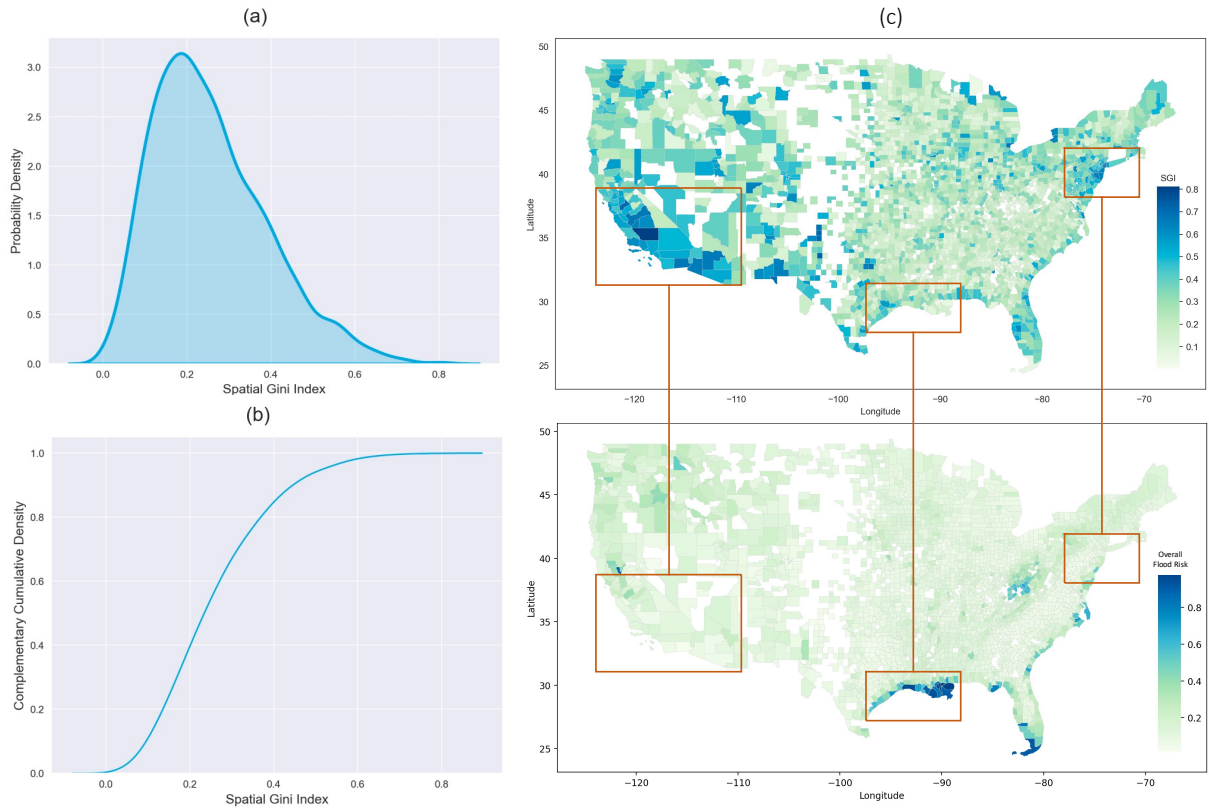

**Fig 1** Disparities of spatial inequality in property flood risk among US counties. (a) Probability density function of SGI. (b) Complementary cumulative distribution function of SGI. (c) County-level visualized comparison between the SGI and overall flood risk. The overall flood risk was calculated by dividing the number of properties with flood factor risk scores larger than 5 by the total number of properties in each county (see Methods for detail). The original data comes from First Street Foundation. We focus on 2567 counties in the continental United States for which all data are available.

To further investigate the variations in spatial inequality of property flood risk across US counties, we classified all the counties into different groups: coastline/non-coastline counties, metropolitan/micropolitan/other counties, Asian/Black/White counties, and income Q1/Q2/Q3/Q4 counties. As shown in Fig 2, the results show significant variations in the SGI across different groups, with some groups exhibiting significantly higher levels of spatial inequality in property flood risk than others. For instance, the boxplot for coastline versus non-coastline counties indicates that the coastline counties tend to have higher SGI, suggesting a higher degree of spatial inequality in property flood risk within these areas. Similarly, the metropolitan counties exhibit higher SGI than micropolitan counties and other counties, indicating that metropolitan counties have a more uneven distribution of property flood risk. The spatial visualizations of SGI distribution in these groups can be found in Fig S1 of Supplementary Information. These results highlight the importance of considering differences in urban form and structure that shape the spatial inequality of property flood risk of counties.

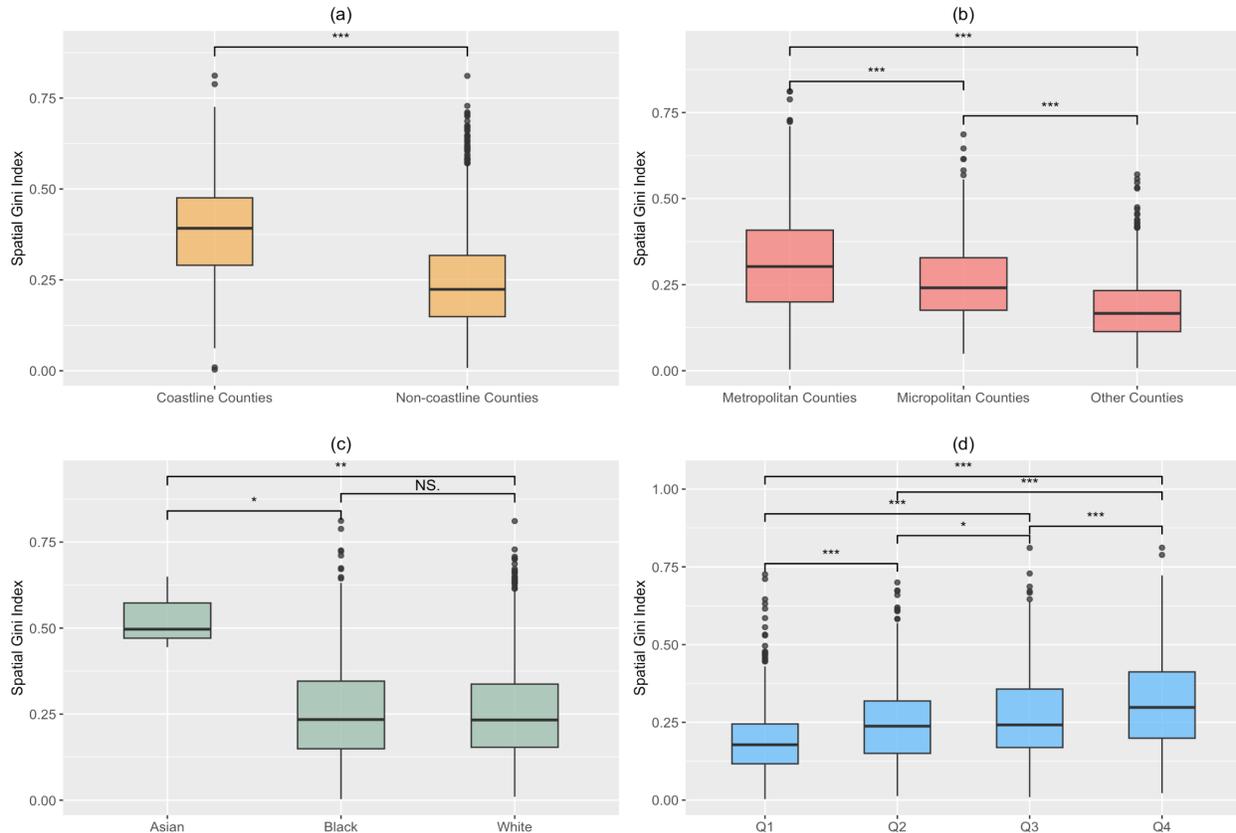

**Fig 2** Variations in spatial inequality of property flood risk among county groups. (a) Coastline/non-coastline counties, as defined by US Census Bureau [27]. We identified 211 coastline counties and 2356 non-coastline counties. (b) Metropolitan/micropolitan/other counties, as identified by US Census Bureau [28]. The numbers are 1101, 601, and 865, respectively. (c) Asian/Black/White counties. The overrepresented race for each county is identified by comparing the proportion of non-Hispanic White, non-Hispanic Black, and non-Hispanic Asian populations to the average for all the counties. We totally have 2025 White-dominated counties, 539 Black-dominated counties, and 3 Asian-dominated counties. We obtained the racial population data from the 2020 race and ethnicity data from US Census Bureau [29]. (d) We used the median income quantile from 2020 American Community Survey [30] to distinguish counties groups: Q1 (642 counties), Q2 (641 counties), Q3 (642 counties), and Q4 (642 counties). Note: ***$p \leq 0.01$, **$p \leq 0.05$, * $p \leq 0.1$. P-values are obtained from t-tests.

**Empirical statistics of urban form and structure features**

We collected a diverse range of datasets, enabling us to capture various heterogenous features related to urban form and urban structure. In terms of urban form, we measured minority segregation, income segregation, population density, and gross domestic product (GDP), while for urban structure, we focused on urban centrality index (UCI), point of interest (POI) density, road density, and human mobility index (HMI). The definition of features and the datasets used to develop these features can be found in the Methods section.

We divided these features into two aspects as they represent different dimensions of the urban environment that could impact the spatial inequality of property flood risk. Urban form captures social and economic factors that influence the spatial distribution of populations and concentration of economic activities. Urban structure, on the other hand, relates to the structural layout of cities, which can affect land use and development patterns shaping property flood risks. These features could potentially explain the spatial inequality of property flood risk. For example, high levels of minority segregation could be associated with redlining that exacerbates property flood risk in already vulnerable areas, while dense urban centers with a high concentration of POI and human mobility could exacerbate impervious surface and reduce green space.

Our initial analysis involved mapping the eight features to examine variations in terms of urban form and structure among counties. Fig 3 illustrates the distribution for the eight features, revealing that the heterogeneity of urban form and structure among US counties. With the exception of road density and income segregation, all features showed higher values in coastal areas compared to non-coastal areas. Counties in metropolitan areas, such as those in California, Florida, and the Northeastern metropolitan area exhibited particularly high values in the features of population density, POI density, minority segregation, UCI, GDP, and HMI.

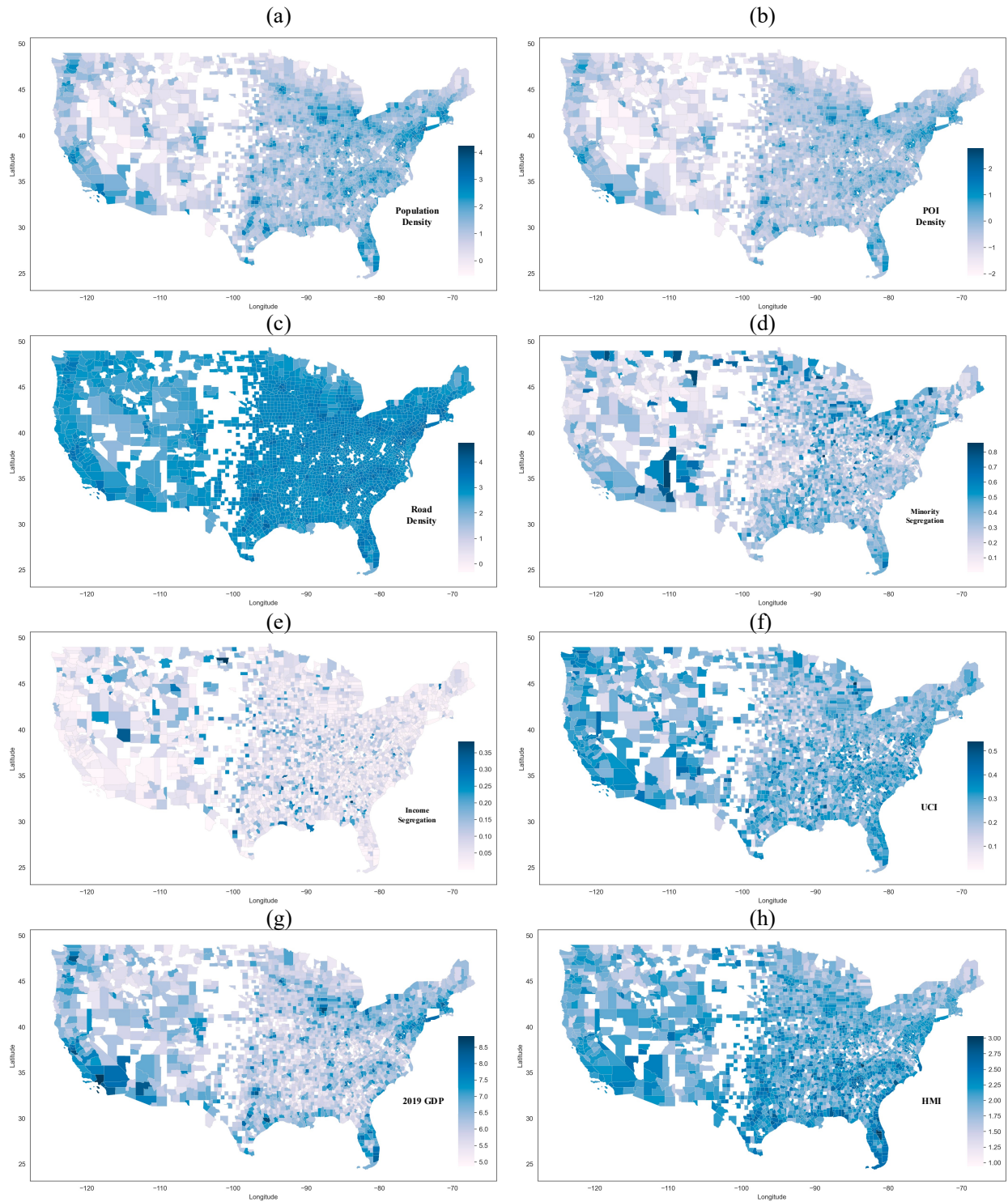

**Fig 3** Heterogeneity in urban form and structure features among US counties. (a) Population density. (b) POI density. (c) Road density. (d) Minority segregation. (e) Income segregation. (f) UCI. (g) 2019 GDP. (h) HMI. Since population density, POI density, road density, and 2019 GDP have a large scale, we used logarithmic transformation of values. See Methods for metrics definition and data processing.

In the next step, we examined the urban form and urban structure features for counties with different levels of spatial inequality of property flood risk. Fig 4 illustrates that there are significant differences in the features of urban form and urban structure among counties with different levels of spatial inequality in property flood risk. Most features exhibit a positive relationship with the extent of spatial inequality in property flood risk, indicating that higher levels of spatial inequality in property flood risk are associated with a greater extent of these features. Notably, population density, POI density, and GDP show the most significant relationship with spatial inequality of property flood risk. This result suggests that counties with greater population density, POI density, and GDP have a greater spatial inequality in property flood risk. Higher population density may lead to a greater concentration of people and property in flood-prone areas, increasing the disparity in property flood risk. Similarly, higher levels of POI density and GDP may indicate greater economic activity and development, which could be associated with denser development that a greater variability in property flood risk. Our observation implies that inequality cannot be simply quantified using only one urban feature due to the complex mechanisms and interaction of urban features. Hidden correlation and interactive pathways of the urban form and structure features causing in inequalities exist and remain underexplored without relying upon further methods. In addition, we ranked the top nine counties in the level five of spatial inequality in property flood risk and showed their statistics for the eight urban form and structure features. (See Fig S2 in Supplementary Information.)

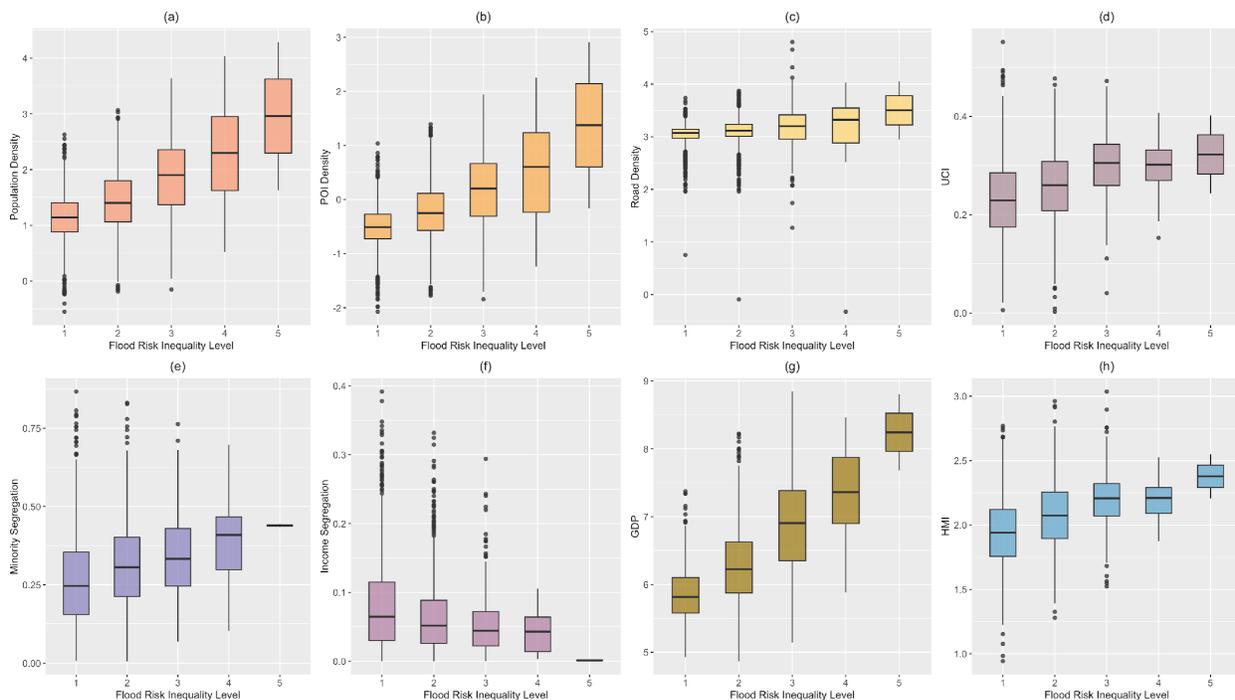

**Fig 4** Disparities of urban form and urban structure features in different levels of spatial inequality of property flood risk. (a) Population density. (b) POI density. (c) Road density. (d) UCI. (e) Minority segregation. (f) Income segregation. (g) 2019 GDP. (h) HMI. Spatial inequality of property flood risk was categorized into five levels: 0 to ≤20% (level 1, minor inequality), 20% to ≤40% (level 2, moderate inequality), 40% to ≤60% (level 3, major inequality), 60% to ≤80% (level 4, severe inequality), and 80% to ≤100% (level 5, extreme inequality).

**Correlation between urban characteristics and spatial inequality of property flood risk**

Next, we analyzed the correlation between the eight urban form and structure features and spatial inequality of property flood risk. The results shown in Fig 5 suggest that seven features were positively correlated with spatial inequality of property flood risk, while income segregation was negatively correlated. The result is quite similar to the boxplots shown in Fig 4. We also calculated the Kendall coefficient, Pearson coefficient, and Spearman coefficient for each feature to further explore the statistical significance of the correlation. (See Fig S3 in Supplementary Information.) The detail related to the regression model and statistical test can be found in Methods.

Taking GDP as an example, the distributions of GDP and the spatial inequality of property flood risk measured by SGI are approximately normal, with rugs shown in Fig 5 (g). The Kendall rank correlation reaches 0.428, the Spearman rank correlation reaches 0.602, and the Pearson correlation coefficient approaches 0.641. All measures are statistically significant with p<0.001, indicating a strong positive correlation between the GDP and the extent of spatial inequality in property flood risk. Other features such as UCI and population density show a similar trend. The strong correlation between GDP, UCI, population density, and SGI servers as an important indication of the role greater economic activities, denser development, and more centralization of facilities play in spatial inequality of property flood risk in cities. That is, a pronounced concentration of economic activities, population density, and facility centralization shape spatial inequality of property flood risk in the United States. The result related to income segregation reveals a reverse relationship compared with the other features. The Kendall rank correlation between income segregation and SGI reaches −0.113, the Spearman rank correlation reaches −0.169, and the Pearson correlation coefficient approaches −0.207. These significant measures signify a moderate negative correlation between income segregation and spatial inequality of property flood risk. This result suggests that counties with a greater income segregation have a lesser extent of spatial inequality in property flood risks.

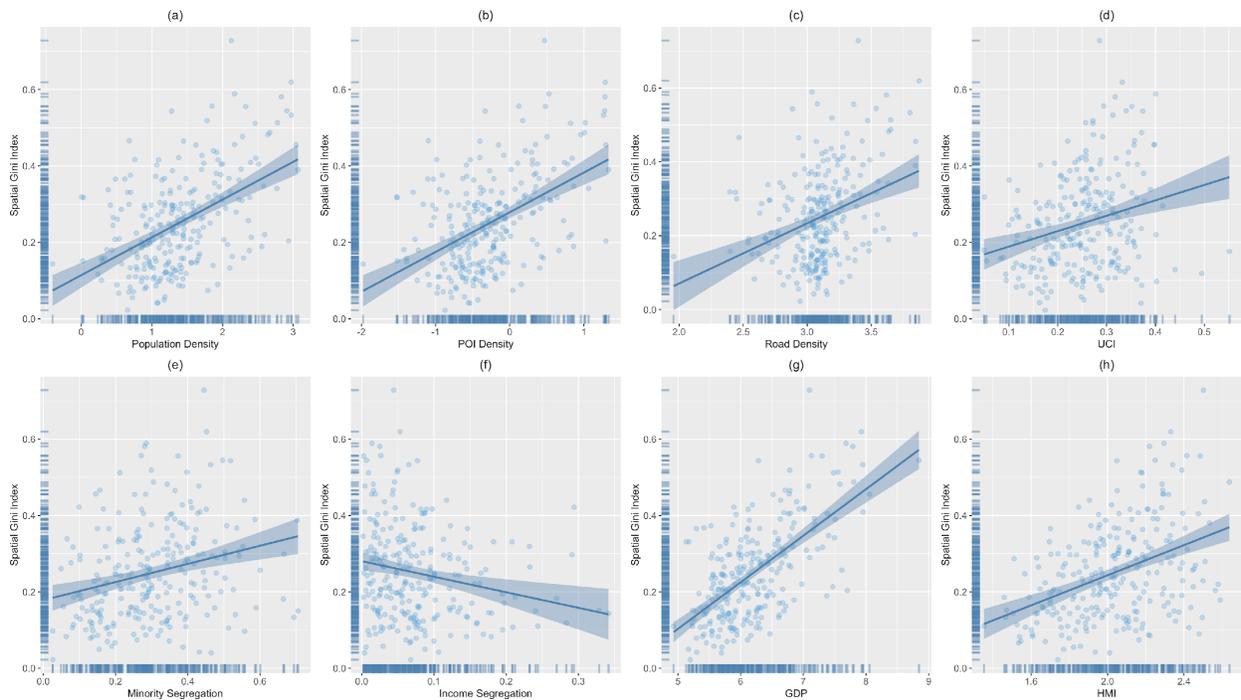

**Fig 5** Correlation between urban form and structure features and spatial inequality of property flood risk. (a) Population density. (b) POI density. (c) Road density. (d) UCI. (e) Minority segregation. (f) Income segregation. (g) GDP. (h) HMI. The variable distribution was given by rugs. The steel blue points show the distribution of 10% raw data. We used the ordinary least squares (OLS) regression model to fit the data and the confidence interval was set to be 90%.

**Pathways to spatial inequality of property flood risk among US counties**

In the next step, we first implemented principal component analysis (PCA), a statistical technique used for dimensionality reduction [31], to the eight features to identify the most important components of urban form and structure that contribute to the spatial inequality of property flood risk. The PCA result is shown in Fig S4 of Supplementary Information. The best number of principal components was selected as three, and the cumulative explained variance of 90.59% indicates that these three principal components capture a significant amount of the variability in the original data and provide a meaningful representation of the urban form and structure.

We defined the three principal components in Table 1. The first principal component is named development density, which includes the features of population density, POI density, and road density, explaining 33.41% of the total variance. This component represents the level of urbanization and built environment density in a given area. The second component is defined as centrality and segregation, explaining 27.56% of the total variance and including the features of UCI, minority segregation, and income segregation. This component represents the level of social and economic segregation, as well as the degree of urban centralization in a given area. The third component, economic activity, explains 29.62% of the total variance and includes the features of GDP and HMI. This component represents the level of economic activity and mobility in a given area.

| Table 1 Definition of the three principal components and their composed features. | | | |
|---|---|---|---|
| Principal component | Name | Proportion of explained variance | Main features |
| PC1 | Development Density (DD) | 33.41% | Population density<br>POI density<br>Road density |
| PC2 | Centrality and Segregation (CS) | 27.56% | UCI<br>Minority segregation<br>Income segregation |
| PC3 | Economic Activity (EA) | 29.62% | GDP<br>HMI |

Upon identification and labeling of the principal components, an entropy-based classification and regression tree model (CART) was implemented to identify pathways that could lead to different levels of spatial inequality of property flood risk among counties by involving the three principal components as the predictor variables and SGI levels as the response variable. (See Methods for detail.)

Based on the model training approaches, which involved normalizing the three principal components to a [0, 1] range, performing an 80:20 train-test split, and implementing 10-fold cross-validation, the model achieved strong performance. Specifically, we measured the model's accuracy score (0.8284 for training data and 0.8178 for testing data), precision, recall, and F1 score (see Table S1 in Supplementary Information). The results show that the model performs well in accurately predicting spatial inequality of property flood risk in counties based on the pathways with combinations of principal components. In order to show as many pathways as possible and avoid overfitting at the same time, we set the minimum split leaf size to 100 counties and limit the tree depth to 7. The tree graph was generated in Fig 6 (a).

In total, we obtained 14 pathways with a single strong majority category. Some tree leaves were combined because of the same pathway structure in Fig 6 (a). The spatial visualization of the counties in the 14 pathways is shown in Fig 7 (a). Each pathway is composed of no more than 200 counties. We eventually extracted the 14 pathways through the structure of the decision tree, some of which are shown in Fig 6 (b). All the pathways can be found in Fig S5 of Supplementary Information. Using Pathway 1 as an example, the decision tree classified 597 counties as having a minor level of spatial inequality of property flood risk.

The top leaf of the tree split the principal component of development density to less than 0.442, while the second leaf split the economic activity to less than 0.432. Development density was again split in the third leaf to less than 0.296. Combining these three leaves, we derived the eventual range for Pathway 1 as [0, 0.296) for development density and [0, 0.432) for economic activity.

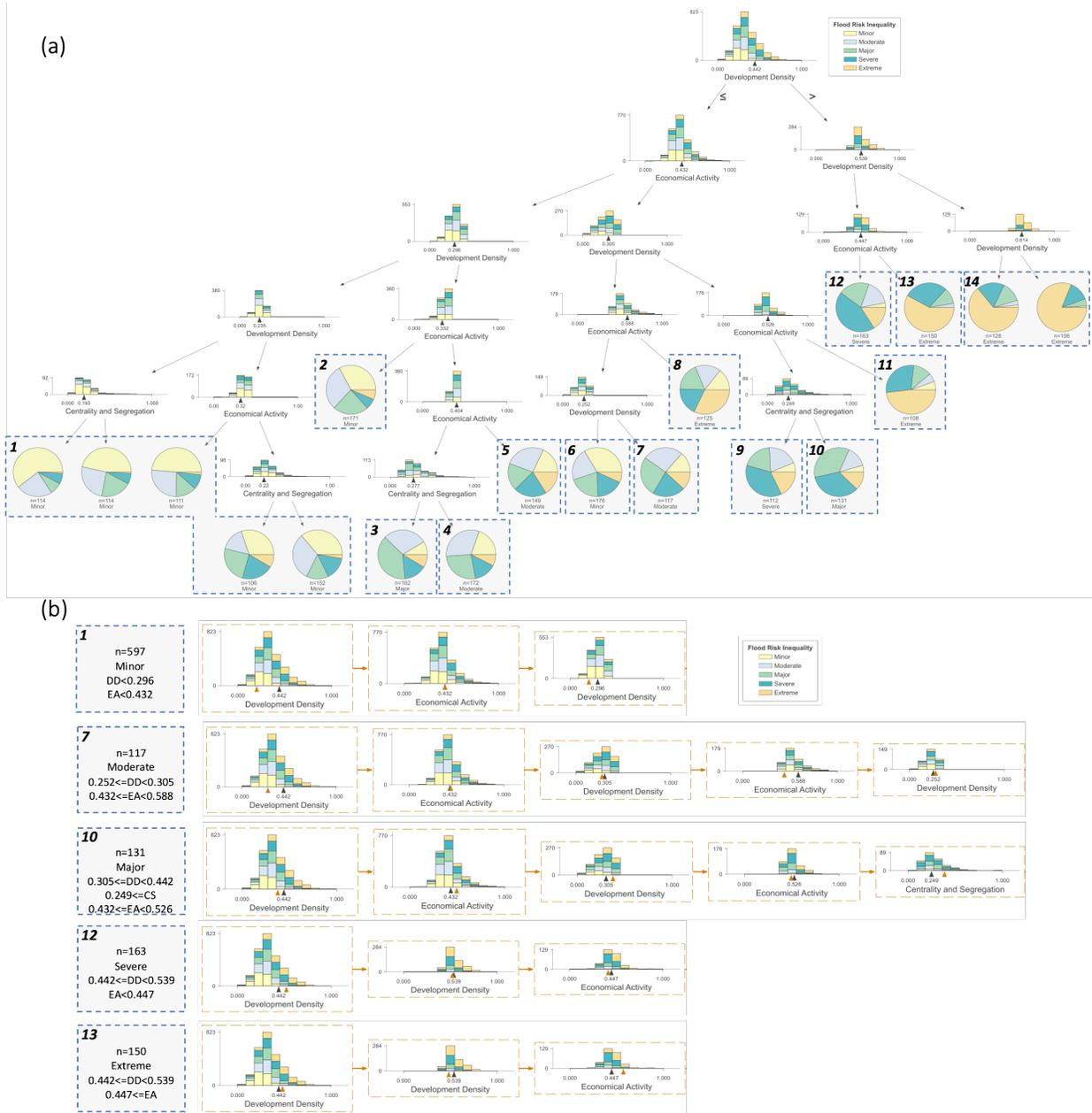

**Fig 6** CART decision tree in terms of different levels of spatial inequality in property flood risk. (a) The completed CART decision tree with pathways marked in the terminal leaves. (b) Example pathways for each level of spatial inequality in property flood risk. The histogram shows the feature space distribution of the five levels of spatial inequality in property flood risk. We use a stacked histogram so that overlap is clear in the feature space between samples with different target classes. The feature space of a left child is everything to the left of the parent's split point in the same feature space, similarly for the right child. The height in the Y axis of the stacked histogram is the total number of samples from all classes and multiple class counts are stacked on top of each other. The black wedge highlights the split point and identifies the exact split value. The orange wedge highlights the split direction. The category with the largest area in the pie chart is considered to be the result of the classification prediction. "n" represents the number of counties in the pathway. EA, DD, and CS represent economic activity, development density, and centrality and segregation, respectively.

To classify the 14 pathways based on their levels of spatial inequality in property flood risk, three pathways were identified as minor inequality in the most significant predicted outcomes, while three were classified as moderate, two as major, two as severe and four as extreme. For pathways with the same level of spatial inequality in property flood risk, we synthesized them and determined the range for the principal components (Fig 7 (c)). We also created a spatial visualization of the outcomes (Fig 7 (b)). Results show that the western part of the United States, as well as counties in the Coast Gulf, Florida Peninsula, Northeastern metropolitan areas, and Great Lakes Region, exhibit a higher level of severe and extreme spatial inequality of property flood risk. In contrast, the contiguous counties in the central and eastern areas exhibit relatively low spatial inequality of property flood risk. We also list example counties for all the pathways to spatial inequality of property flood risk, which can be found in the Table S2 of Supplementary Information. For example, New York County in New York State, Los Angeles County in California, Harris County in Texas are on the top of the list for the extreme level of spatial inequality in property flood risk.

Our analysis demonstrates that the principal components of development density and centrality and segregation have a substantial influence on determining the extreme level of spatial inequality in property flood risk. These components exhibit a range of 0 to 1, indicating their consistent impact on the property flood risk, regardless of the specific values they hold. On the other hand, the principal component of economic activity was found to have a range of [0.447, 1]. This observation suggests that economic activity, encompassing GDP and HMI characteristics, plays a crucial role in predicting the extreme level of spatial inequality in property flood risk. This discovery underscores the significance of considering a region's economic activity in evaluating spatial inequality of property flood risk and formulating policies to alleviate the impact of flood on the most vulnerable populations.

The minor level of spatial inequality in property flood risk is another example where we observe a relatively wide range for the three principal components. Specifically, the range for development density is [0, 0.442], which includes features such as population density, POI density, and road density. Meanwhile, the range for economic activity, which includes features such as GDP and HMI, is [0, 0.588]. This suggests that a county can be classified as having a minor level of spatial inequality of property flood risk if it satisfies both ranges for development density and economic activity. It is also important to acknowledge that centrality and segregation play a highly significant role in this classification, as their range remains consistent across all counties, spanning from 0 to 1.

Overall, Fig 7 (c) highlights the complex relationship between different components and their contribution to the spatial inequality of property flood risk. By identifying the specific ranges for each pathway, we can better understand the factors that contribute to the spatial inequality of property flood risk and inform policy-making for more targeted and effective flood risk management.

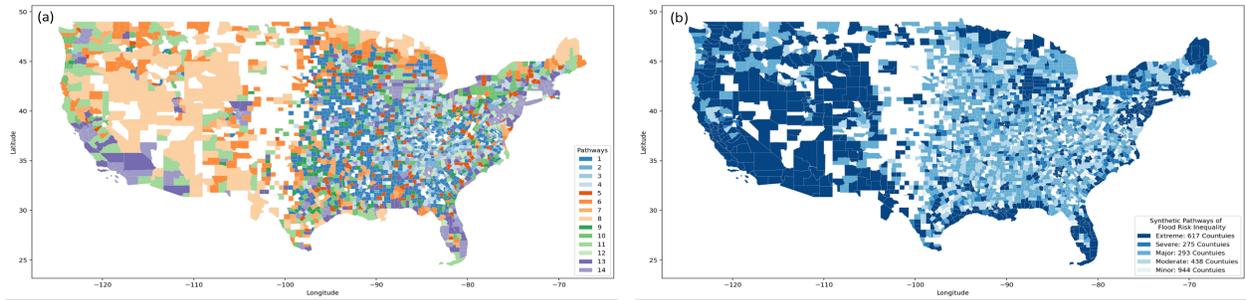

**Fig 7** Spatial visualization of the pathways and their range regrading to the principal components. (a) Spatial distribution for the original fourteen pathways. (b) Spatial distribution for the five synthetic pathways. (c) Range values of the principal components in the five synthetic pathways.

**Discussion**

This study explores the relationship between urban form and structure and spatial inequality of property flood risk using rich datasets across the United States. As cities continue to grow and develop, it is essential to understand ways in which features of urban form and structure could shape spatial inequality in flood risk to properties and people. Such understanding is particularly important for devising integrated urban design strategies to address flood risk in conjunction with urban growth and development plans.

The findings from this study advance our understanding of the complex interplay between urban form and structure and spatial inequality of property flood risk in multiple important aspects. First, the findings provide empirical evidence for the presence of significant property flood risk inequality in metropolitan and coastal counties in the United States. The results indicate that the western part of the country, as well as the counties in the Coast Gulf, Florida Peninsula, the Northeastern metropolitan areas, and the Great Lakes Region, exhibit greater levels of spatial inequality in property flood risk. For example, New York County in New York State, Los Angeles County in California, Harris County in Texas are on the top of the list for the extreme level of spatial inequality in property flood risk.

Second, the findings reveal the principal component factors and pathways related to urban form and structure that shape spatial inequality of property flood risk. Of particular importance is the identification of development density as the prominent principal component factors in explaining variations in spatial inequality of property flood risks. These findings imply that urban growth and development strategies that exacerbate development density in cities could yield significant spatial inequality of property flood risk. Increasing development density in cities exacerbates the existing flood risk hotspots [32] and yields new hotspots [33] all of which would increase spatial inequality of property flood risks. Cities with already significant development density could primarily address spatial inequality of property flood risk by alleviating development density. Urban development strategies such as zoning regulations [34], mix-use development [35], and prioritization of open area [36] that are shown to alleviate development density in cities could also help address spatial inequality of property flood risk.

Third, the findings also provide data-driven insights regarding the tradeoff between economic development and flood risk in cities. The results of pathway analysis show that cities with moderate development density would have severe or extreme spatial inequality of property flood risk if the level of economic activity is high. The implication of this finding is that, in order for cities to pursue economic development and growth while addressing property flood risk and its spatial inequality, it is important to focus on controlling development density. The combination of dense development and high economic activity would be a pathway for severe or extreme property flood risk in cities.

Fourth, these findings underscore the importance of integrated urban design strategies to address complex urban issues at the intersection of development, growth, and flood risk management. Typically, urban issues are addressed separately using isolated plans and policies. This study shows that urban form and structure which is shaped by various urban growth and development features shape the spatial inequality of property flood risk. Hence, it is important for various urban plans and policies to be integrated to identify strategies that address different vexing urban issues simultaneously. Also, this study shows the importance of data-driven methods for various fields of urban planning, engineering, city science, geography, and environmental sciences in devising integrated urban design strategies based on examining heterogenous urban features and their interactions to better understand ways different urban features (individually and collectively) shape different outcomes related to sustainability, flood risk, and other urban phenomena.

The contributions of this study inform researchers and practitioners in various fields including urban planning, engineering, city science, and flood risk management about the interplay between urban form and structure features and spatial inequality of property flood risk and also show new avenues for future research directions. For example, based on the findings of this study, future research can investigate causal relationships between features of urban form and structure and property flood risks in cities. For example, causal inference techniques based on spatial deep learning can be adopted. Using such models, future scenarios of urban growth and development and their effects on property flood risk and its spatial heterogeneity could be investigated.

## Methods

### Definition and data for spatial inequality of property flood risk

Spatial inequality of property flood risk refers to the uneven distribution of flood risk across different geographic areas. It reflects the degree to which some areas have a higher likelihood of experiencing flood damage to properties than others, often due to differences in physical, environmental, and socio-economic factors.

The raw dataset of property flood risk was obtained from Flood Factor, a model created by First Street Foundation [37]. The Flood Factor model assesses the flood risk of every property in an area and assigns it a risk score from 1 to 10. A score of 1 indicates a low chance of flooding within the next 30 years; a score of 10 indicates a high chance of flooding. The scores are derived from publicly available and third-party data and take into account elevation, precipitation, environmental changes over time, and community protection projects [37].

Spatial Gini index (SGI) is a measure of spatial inequality. It is calculated as the area difference between a perfectly equal distribution and the actual distribution [38, 39]. An SGI of 0 represents a perfectly equal distribution of the feature of interest and 1 would describe a distribution where different areas have different values of the feature of interest [38]. In this study, SGI captures the spatial heterogeneity of property flood risk in a county. We calculated the percentage of property flood risk for each county by dividing the number of properties with Flood Factor score larger than 6 to the total number of properties, denoted here by $x_i$. The entropy-based SGI is given by [40]:

$$SGI = \frac{\sum_{i=1}^{N}\sum_{j=1}^{N} w_{ij}|x_i - x_j| + (1 - w_{ij})|x_i - x_j|}{2n^2\langle x \rangle} \quad (1)$$

where $N$ is the number of neighborhoods, and $\langle x \rangle = \frac{1}{N}\sum_i x_i$ is the mean of the variable of interest. The spatial weight $w_{ij}$ is defined according to the adjacency matrix $A$ where $w_{ij} = 1$ if two areas are neighbours, and 0 otherwise. The diagonal elements $w_{ii} = 0$ as defined in $A$ and $W$ corresponds to the sum of all weights.

### Definition and data for urban form features

**GDP:** To estimate the status of the economic development of the county, we adopted the 2019 data of gross domestic product for each county. The data are provided by the Bureau of Economic Analysis in the US Department of Commerce [41].

**Population density:** The population size was obtained from the 2020 race and ethnicity data from US Census Bureau [29]. We calculated the population density at the county level by dividing the total population of the county by its land area. Land area data was also obtained from US Census Bureau [42].

**Minority segregation and income segregation:** Urban segregation refers to the physical and social separation of different racial, ethnic, and socioeconomic groups within a city [43]. This separation can take many forms, including minority segregation and income segregation. One of the key consequences of urban segregation is that it often leads to unequal distribution of resources, as well as increased exposure to environmental hazards such as flooding [44, 45].

In this study, we adopted the Dissimilarity Index (DI) to evaluate minority segregation and income segregation. The DI is a measure of spatial segregation that indicates the extent to which two groups are evenly distributed across different areas, which ranges from 0 (indicating perfect evenness) to 1 (indicating complete separation) [46, 47]. We calculated the DI based on the proportion of minority population (for minority segregation) and the proportion of low-income population (for income segregation) at the census-tract level relative to the county level [48]:

$$DI = \frac{1}{2} \sum_{i=1}^{n} \left| \frac{x_i}{X} - \frac{y_i}{Y} \right| \qquad (2)$$

where $x_i$ is the minority population (or low-income population) in the smaller geographical unit; $X$ is the the minority population (or low-income population) in the larger geographical unit. $y_i$ is the reference population in the smaller geographical unit; $Y$ is the reference population in the larger geographical unit.

For minority segregation, we collected the racial population data from the 2020 race and ethnicity data from US Census Bureau [29]. The primary racial groups in this study are non-Hispanic White, non-Hispanic Black, and non- Hispanic Asian residents. We considered the non-Hispanic Black, and non- Hispanic Asian as the minority population and non-Hispanic White as the reference population.

For income segregation, we extracted median income data from the 2020 American Community Survey (ACS) [30]. This study used the 5-year estimates of median income due to the broader coverage of areas, larger sample size, and higher precision, making the data more reliable than 1-year and 3-year estimates. We used the quantile income groups of a county (Q1 to Q4) to indicate income levels with Q1/Q2 representing low-income groups and Q3/Q4 represent high-income groups, respectively.

**Definition and data for urban structure features**

**POI density:** To capture the distribution of facilities, we adopted the 6.5 million active POI data in the US from SafeGraph [49]. The dataset includes basic information about POIs, such as POI IDs, location names, geographical coordinates, addresses, brands, and North American Industry Classification System (NAICS) codes to categorize POIs. The NAICS code is the standard used by federal statistical agencies in classifying business establishments [50]. In this study, we selected 10 essential types of POIs that are closely relevant to human daily lives: restaurants, schools, grocery stores, churches, gas stations, pharmacies and drug stores, banks, hospitals, parks, and shopping malls. We counted the number of POIs in each county and calculated its density as their facility distribution feature.

**Road density:** To capture the distribution of transportation network, we extracted data from Open Street Map [51] to calculate the density of road segments in counties. We estimated complete road networks from the raw data by assembling road segments. Since the lengths of road segments created by the source were in close proximity, we calculated road density by dividing the number of road segments by the areas of a county.

**Urban centrality index:** We adopted urban centrality index (UCI) to characterize the centralization degree of the facilities in a county. UCI is the product of the local coefficient and the proximity index [52]. The local coefficient was computed based on the number of POIs within each census tract; the proximity index was computed based on the number of POIs within each census tract along with a distance matrix that considered the distance between census tracts. The value of UCI ranges from 0 to 1. The values close to 0 indicate polycentric distribution of facilities within a county, while the values close to 1 indicate monocentric distribution of facilities. The indices are formulated as follows [52]:

$$LC = \frac{1}{2} \sum_{i=1}^{N} (k_i - \frac{1}{N}) \qquad (3)$$

$$PI = 1 - \frac{V}{V_{max}} \quad (4)$$

$$V = K' \times D \times K \quad (5)$$

where $N$ is the total number of census tracts in a county; $K$ is a vector of the number of POIs in each census tract; $k_i$ is a component of the vector $K$; $D$ is the distance matrix between census tracts; $V_{max}$ is calculated by assuming that the total POIs are uniformly settling on the boundary of the county; $LC$ is the local coefficient, which measures the unequal distribution; $PI$ is the proximity index, which resolves the normalization issue; $V$ is the Venables Index.

**Human mobility index:** To understand the inequality of population activities, we employed mobile phone data from Spectus Inc. to develop the metric of human mobility index (HMI). The data has a wide set of attributes, including anonymized user ID, latitude, longitude, POI ID, time of observation, and the dwelling time of each visit [53]. Prior studies found that Spectus mobile phone data is representative to describe human activities and mobility [54-56]. Hence, the feature generated using the dataset should be representative and valid for our analyses. We extracted the data from April 2019 (28 days) to account for the variation of population activities on weekdays and weekends. Our period is also during regular conditions when no external extreme events perturbed human activities. To develop the HMI, we first assigned each visit point $v_i$ to a defined CBG in a county. Then, we calculated HMI as follows:

$$HMI = \frac{\sum_{i=1}^{n} v_i}{28n} \quad (6)$$

where $n$ denotes the number of CBGs in a county.

We finally mapped the values of HMI to the range from 0 to 1 using min-max scaling. The proximity of HMI values to 0 or 1 indicates the level of human mobility and activity, with values closer to 0 indicating lower activity and values closer to 1 indicating higher activity in a county.

**Statistical analysis**

**Ordinary least squares regression model:** We employed an ordinary least squares regression model to capture the relationships between urban form and structure and spatial inequality of property flood risk among counties and to understand the relative importance of each feature [57]:

$$y_i \sim \beta_0 + \beta_1 x_{i,1} + \beta_2 x_{i,2} + \beta_3 x_{i,3} + \beta_4 x_{i,4} + \beta_5 x_{i,5} + \beta_6 x_{i,6} + \beta_7 x_{i,7} + \beta_8 x_{i,8} + \varepsilon_i \quad (7)$$

where, $y_i$ is the SGI of county $i$; $x_{i,1}$ to $x_{i,8}$ are the features of urban form and structure; $\beta$ are coefficients; $\varepsilon_i$ is the error term.

In the regression, since the values of POI density, population density, road density, and GDP have a much larger scale than other variables, we used logarithmic transformation of values. Three statistical tests, Kendall's tau test, Pearson's correlation test, and Spearman's rank correlation test, were then conducted for the correlation analyses to examine statistical significance and determine feature importance.

**Classification and regression tree model:** The classification and regression tree (CART) model is an unsupervised machine learning algorithm used to build a decision tree by recursively splitting the data based on the predictor variables to minimize the entropy in the response variable [58]. The decision tree consists of a series of nodes, each representing a split in the data based on a particular predictor variable, and terminal nodes representing the predicted response variable for a given combination of predictor variable values. The method to identify the best splits is to minimize the entropy. If the entropy of the two child nodes is not lower than that of a parent node, splitting will be terminated any further. The entropy ($E$) is given by [59]:

$$E = -\sum_{i=1}^{n} p_i log_2(p_i) \tag{8}$$

where $p_i$ is the fraction of items in the class $i$.

In this study, we categorized the SGI into five levels: 0 to ≤20% (minor inequality), 20% to ≤40% (moderate inequality), 40% to ≤60% (major inequality), 60% to ≤80% (severe inequality), and 80% to ≤100% (extreme inequality). Then, we implemented a CART classification algorithm using the principal components as the predictor variables and SGI levels as the response variable.

Decision trees were utilized to pinpoint the factors that shape pathways to different levels of spatial inequality of property flood risk among different counties. By analyzing the decision trees and the pathways they present, this study aims to shed light on the contributing factors to the spatial inequality of property flood risk in the United States. Our primary objective is to uncover as many pathways as possible while maintaining good performance, considering the balance between complexity and performance. To achieve this, we set the minimum split leaf size to 100 counties and limit the tree depth to 7, enabling us to generate more pathways while also limiting the tree depth to prevent overfitting.

**Data availability**

All data were collected through a CCPA- and GDPR-compliant framework and utilized for research purposes. The datasets of POI, human mobility, and Flood Factor scores that support the findings of this study are available from SafeGraph Inc., Spectus Inc., and First Street Foundation respectively, but restrictions apply to the availability of these datasets, which were used under license for the current study. The datasets can be accessed upon request submitted on safegraph.com, spectus.ai, and firststreet.org, respectively. Other data used in this study are all publicly available.

**Code availability**

The code that supports the findings of this study is available from the corresponding author upon request.

**Acknowledgements**

This material is based in part upon work supported by the National Science Foundation under CRISP 2.0 Type 2 No. 1832662 grant. The authors also would like to acknowledge the data support from SafeGraph, Spectus, and First Street Foundation. Any opinions, findings, conclusions, or recommendations expressed in this material are those of the authors and do not necessarily reflect the views of the National Science Foundation, SafeGraph, Spectus, or First Street Foundation.

**Author contributions**

J.M.: Conceptualization, Methodology, Data curation, Formal analysis, Writing - Original draft. A.M.: Conceptualization, Methodology, Writing - Reviewing and Editing, Supervision, Funding acquisition.

**Competing interests**

The authors declare no competing interests.

**Additional information**

Supplementary information associated with this article can be found after the references.

Supplementary Information for

*Urban Form and Structure Explain Variability in Spatial Inequality of Property Flood Risk among US Counties*

June 02, 2023

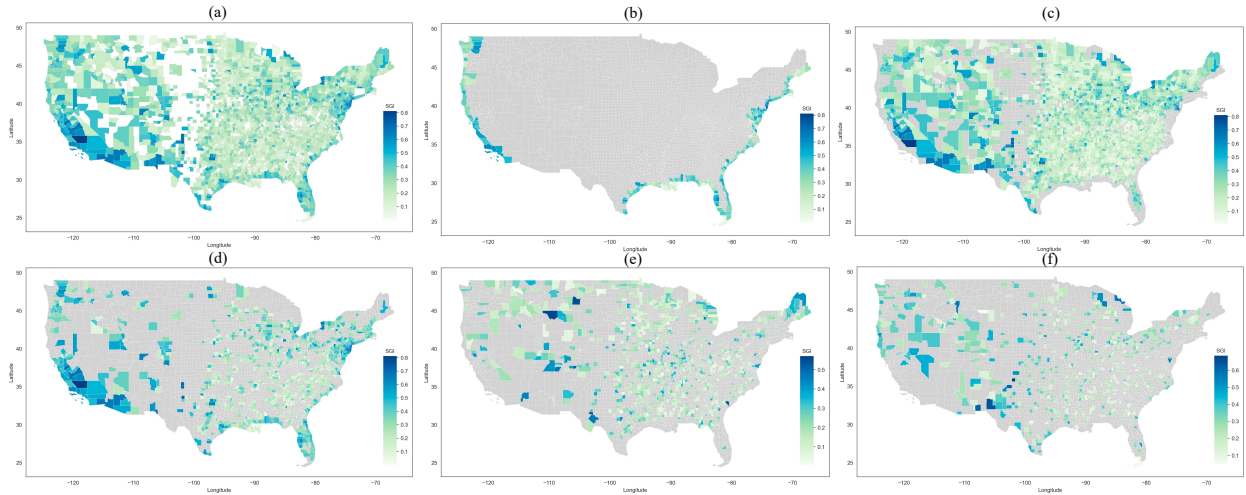

Fig S1 Disparities of spatial inequality in property flood risk among different groups of US counties. We use Spatial Gini Index (SGI) to measure spatial inequality in property flood risk. The results show significant variations in the SGI across different county groups. (a) Spatial distribution of SGI in all of 2567 US counties in our study. (b) Spatial distribution of SGI in US coastline counties. (c) Spatial distribution of SGI in US non-coastline counties. (d) Spatial distribution of SGI in US metropolitan counties. (e) Spatial distribution of SGI in US micropolitan counties. (f) Spatial distribution of SGI in other counties (non-metropolitan and non-micropolitan counties).

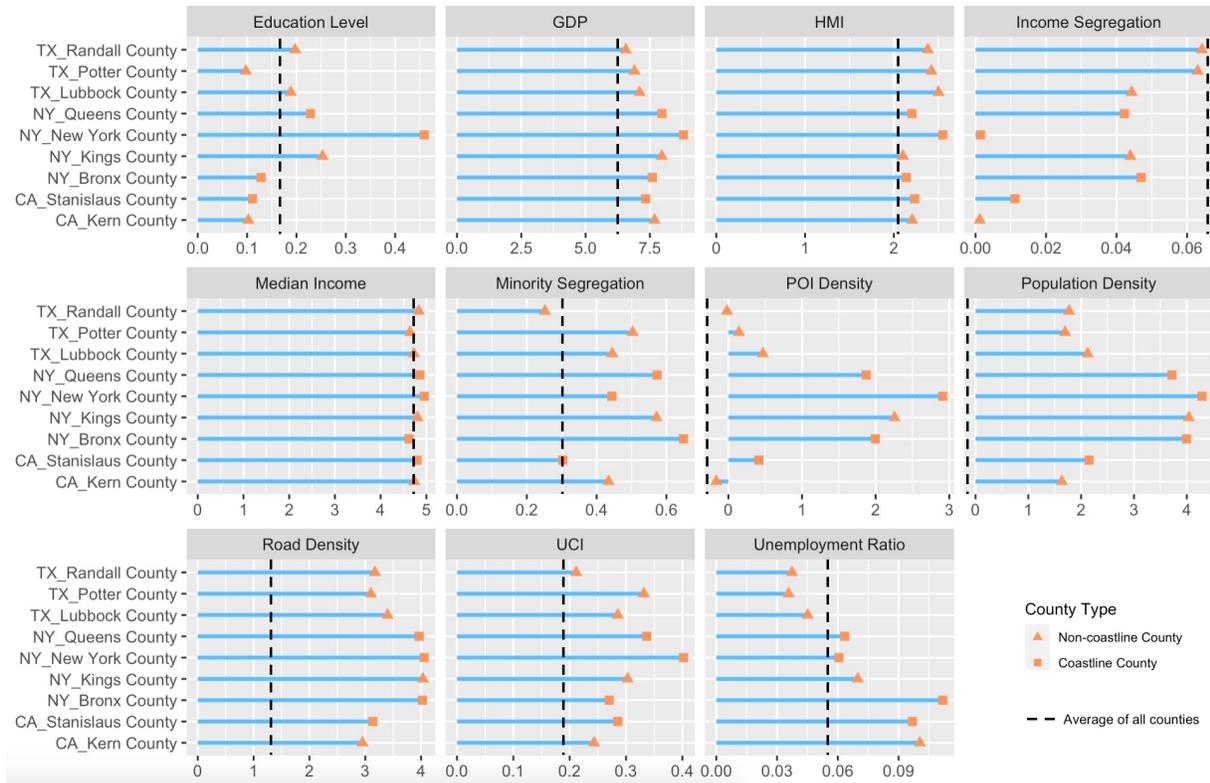

Fig S2 The top nine counties in level five of spatial inequality of property flood risk (extreme inequality). We calculated the values of eight urban form and structure features/education level/median income/unemployment ratio for these counties and compared them to the average of all counties (black vertical dash lines) in our study. We distinguished the coastline counties (square) and non-coastline counties (triangle) with different point shapes. We extracted median household income and total population for different education and employment level at the county level from the 2017 - 2021 American Community Survey 5-year estimates via the US Census Bureau (https://www.census.gov/newsroom/press-kits/2022/acs-5-year.html).

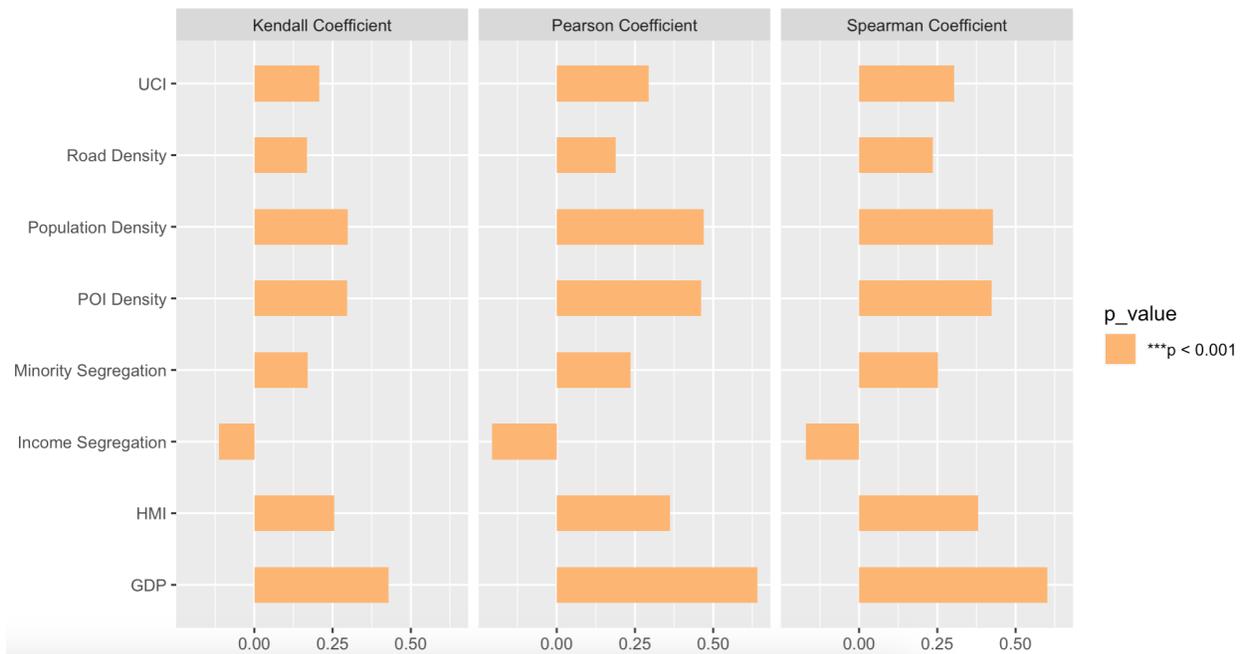

Fig S3 The Kendall coefficient, Pearson coefficient, and Spearman coefficient were calculated and ranked for urban form and structure features. All measures are statistically significant with ***p<0.001.

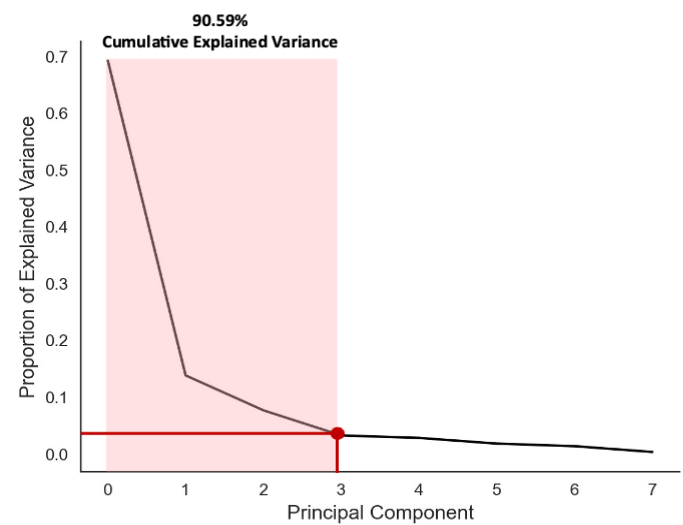

Fig S4 Scree plot for the results of the principal components analysis (PCA). The best number of principal components was selected as three and the cumulative explained variance of 90.59% indicates that these three principal components capture a significant amount of the variability in the original data and provide a meaningful representation of the urban form and structure features.

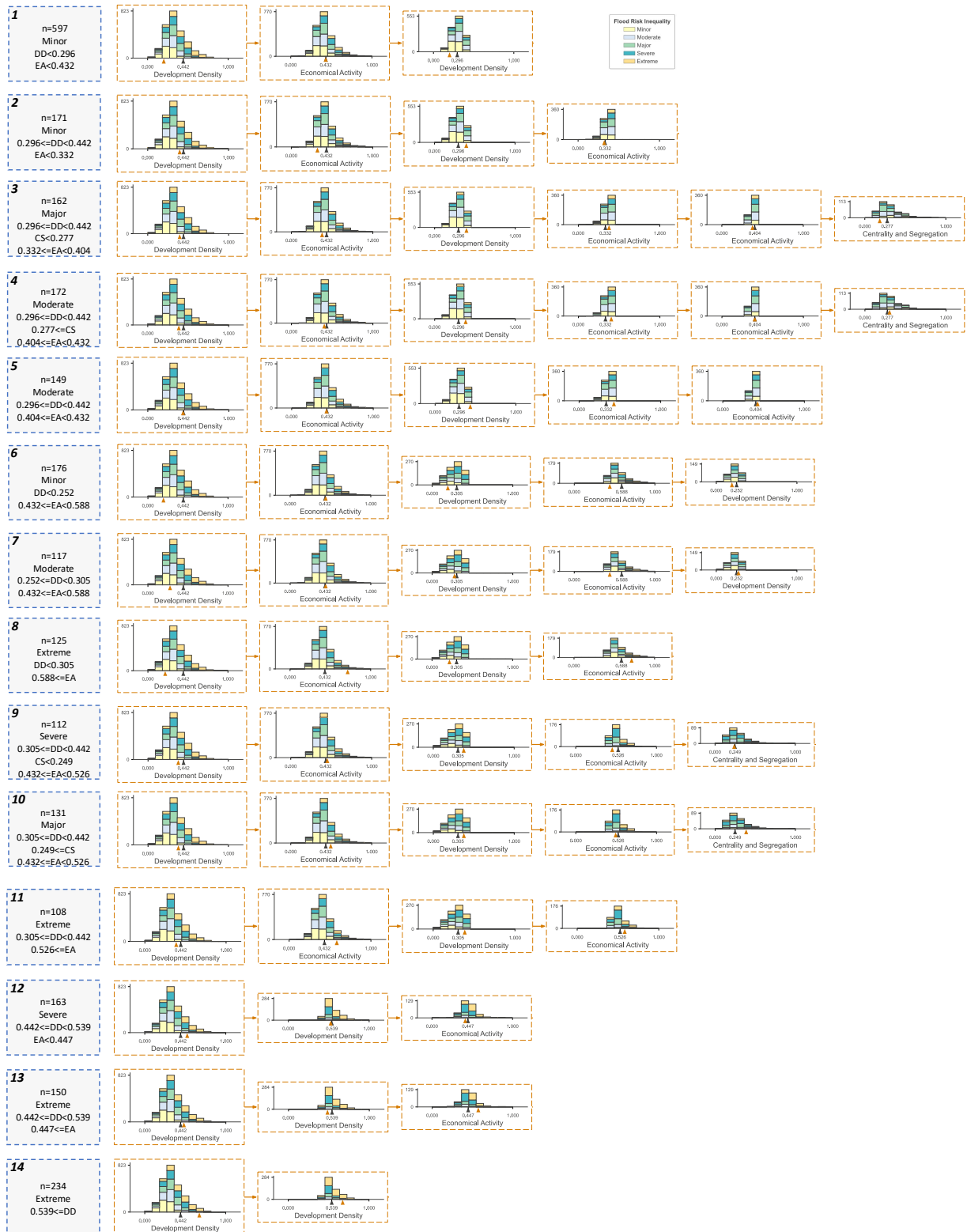

Fig S5 The pathways extracted through the structure of the decision tree show the different levels of spatial inequality of property flood risk. "n" represents the number of counties in the pathway. EA, DD, and CS represent economic activity, development density, and centrality and segregation, respectively. The detailed explanation of the decision tree and pathway can be found in Fig 6 of the main text.

Table S1 The results of CART model performance. We involved normalizing the three principal components to a [0, 1] range, performing an 80:20 train-test split, and implementing 10-fold cross-validation to train the model.

| Performance Statistics | | | | |
|---|---|---|---|---|
| Tree depth | | | 7 | |
| Minimum split leaf size | | | 100 | |
| No. of leaves | | | 19 | |
| **Evaluation on Training Data** | | | | |
| Accuracy Score | | | 0.8284 | |
| Class | Precision | Recall | F1-score | Support |
| 1 (Minor) | 0.84 | 0.97 | 0.90 | 814 |
| 2 (Moderate) | 0.76 | 0.64 | 0.85 | 328 |
| 3 (Major) | 0.72 | 0.70 | 0.84 | 264 |
| 4 (Severe) | 0.85 | 0.78 | 0.89 | 215 |
| 5 (Extreme) | 0.85 | 0.81 | 0.91 | 433 |
| Macro avg | 0.80 | 0.65 | 0.68 | 2054 |
| Weighted avg | 0.82 | 0.83 | 0.80 | 2054 |
| **Evaluation on Testing Data** | | | | |
| Accuracy Score | | | 0.8178 | |
| Class | Precision | Recall | F1-score | Support |
| 1 (Minor) | 0.83 | 0.97 | 0.90 | 195 |
| 2 (Moderate) | 0.76 | 0.66 | 0.77 | 76 |
| 3 (Major) | 0.75 | 0.75 | 0.81 | 62 |
| 4 (Severe) | 0.79 | 0.80 | 0.83 | 42 |
| 5 (Extreme) | 0.84 | 0.82 | 0.88 | 138 |
| Macro avg | 0.80 | 0.65 | 0.68 | 513 |
| Weighted avg | 0.82 | 0.83 | 0.80 | 513 |

Table S2 Example counties for all the pathways to spatial inequality in property flood risk.

| Pathway | Spatial Inequality Level of Property Flood Risk | Example County |
|---|---|---|
| 1 | Minor | TX_Rains County<br>WV_Wyoming County<br>GA_Meriwether County |
| 2 | Minor | NY_Yates County<br>VA_Tazewell County<br>GA_Fannin County |
| 6 | Minor | OK_Love County<br>MI_Iron County<br>CO_Logan County |
| 4 | Moderate | NY_Columbia County<br>TX_Caldwell County<br>MN_McLeod County |
| 5 | Moderate | AR_Hot Spring County<br>TX_Liberty County<br>TX_Navarro County |
| 7 | Moderate | CA_Tehama County<br>FL_Hendry County<br>OK_Kingfisher County |
| 3 | Major | TX_Waller County<br>NY_Cattaraugus Count<br>FL_Santa Rosa Countyy |
| 10 | Major | OH_Ashtabula County<br>TX_Hunt County<br>CA_Amador County |
| 9 | Severe | FL_Citrus County<br>TX_Brazos County<br>FL_Hernando County |
| 12 | Severe | SC_Dorchester County<br>TX_Comal County<br>TX_Kaufman County |
| 8 | Extreme | NY_New York County<br>TX_Travis County<br>CA_San Francisco County |
| 11 | Extreme | CA_San Mateo County<br>CA_San Diego County<br>CA_Los Angeles County |
| 13 | Extreme | NY_Bronx County<br>TX_Harris County<br>NY_Rockland County |
| 14 | Extreme | TX_Dallas County<br>NY_Westchester County<br>NY_Kings County |